\documentclass[conference,a4paper]{APSIPA2018}
\usepackage{multirow,cite,url,color}
\usepackage{graphicx}
\usepackage[fleqn]{amsmath}
\usepackage{amssymb}

\begin{document}

\title{Robust Image Identification for Double-Compressed and Resized JPEG Images}

\author{%
\authorblockN{%
Kenta Iida\authorrefmark{1} and
Hitoshi Kiya\authorrefmark{1}
}
\authorblockA{%
\authorrefmark{1}
Tokyo Metropolitan University\\
E-mail: iken729@gmail.com, kiya@tmu.ac.jp}
}

\maketitle
\thispagestyle{empty}

\begin{abstract}
In the case that images are shared via social networking services (SNS) and cloud photo sharing services (CPSS), it is known that the JPEG images uploaded to the services are often re-compressed and resized by the providers.
Because of such a situation, a new image identification scheme for double-compressed JPEG images having different sizes from that of a singled-compressed one is proposed in this paper.
The aim is to detect a single-compressed image that has the same original image as the double-compressed ones, even when the sizes of those compressed images are different.
In the proposed scheme, a feature extracted from only DC coefficients in DCT coefficients is used for the identification.  
The use of  the feature allows us not only to robustly avoid errors caused by double-compression but also to perform the identification for different size images.
The simulation results demonstrate the effectiveness of  the proposed one in terms of the querying performance.
\end{abstract}

\section{Introduction}
\label{sec:intro}
The growing popularity of photo sharing applications on the Internet has opened new perspectives in many research fields, including the emerging area of multimedia forensics.
Those applications include social network services (SNS) like Facebook and cloud photo sharing services (CPSS) like Google photos.
The huge amount of images uploaded to SNS and CPSS are generally stored in a compressed format as JPEG images, after being resized and re-compressed using different compression parameters from those used for the uploaded images\cite{Intro,Intro1,Intro2}.
Due to a such situation, identifying JPEG images which have the same original image has been required.

Several identification schemes and robust image hashing ones have been proposed to consider the relationship between images \cite{id1,id2,id4,conv,dc,fcs2, fcs1,zpid,zpid1,zpid2,cbir,ih1,ih2,itq}.
They have been developed for the various purposes:producing evidence regarding image integrity, robust image retrieval, finding illegally distributed images and so on.
The conventional schemes for identifying images can be broadly classified into two types: compression-method-dependent and compression-method-independent. 
Compression-method-independent schemes include image retrieval and image hashing-based ones \cite{cbir,ih1,ih2,itq}.
These schemes generally extract features from resized or divided images after decoding images, and then the features are converted to other representations.
For instance, ITQ-based scheme\cite{itq} converts Gist descriptors\cite{gist} generated from divided images.
The compression-method-independent schemes have tried not only to identify images having the different sizes but also to consider several noises including errors caused by lossy compression.
However, they sometimes miss detecting slight differences because they mainly aim to retrieve similar images. 

On the other hand, due to the use of robust features against JPEG errors, compression-method-dependent schemes \cite{conv,dc,fcs2,fcs1,zpid,zpid1,zpid2}  have the stronger robustness than the first type ones.
The schemes\cite{conv, dc,fcs2, fcs1} use positive and negative signs of discrete cosine transform (DCT) coefficients, and the schemes\cite{zpid,zpid1,zpid2} focus on the positions in which DCT coefficients have zero values.
However, the identification for images having different sizes can not be performed.
In addition, most of them do not consider the identification between single-compressed images and double-compressed ones.

Due to such situations, our proposed scheme can robustly identify JPEG images double-compressed under various compression conditions, even if the sizes of the images are different.
The identification is carried out with a feature extracted from DC coefficients.
The use of the DC coefficients-based feature allows us not only to avoid errors caused by double-compression but also to achieved that images having different sizes can be identified.
The simulation results demonstrate that the proposed scheme enables to detect slight differences, even if images are very similar.

\section{Preliminaries}
\subsection{JPEG Encoding}
The JPEG standard is the most widely used image compression standard.
The JPEG encoding procedure can be summarized as follows.

\begin{itemize}
  \item[1)]
Perform color transformation from RGB space to $\mathrm{YC_{b}C_{r}}$ space and sub-sample $\mathrm{C_{b}}$ and $\mathrm{C_{r}}$. 
 \item[2)]
Divide an image into non-overlapping consecutive 8$\times$8-blocks.
 \item[3)]
Apply DCT to each block to obtain 8$\times$8 DCT coefficients $\bf S$, after mapping all pixel values in each block  from [0,255] to [-128,127] by subtracting 128 in general.
 \item[4)]
Quantize $\bf S$ using a quantization matrix $\bf Q$.
\item[5)]
Entropy code it using Huffman coding.
\end{itemize}
A DC coefficient $S(0,0)$ in each block is obtained by the following equation, where $I(b_x,b_y)$ represents a level-shifted pixel value at the position $(b_x,b_y)$ in a block.
\begin{equation}
\label{eq:dc}
S(0,0)= \frac{1}{8} \sum_{b_x=0}^7 \sum_{b_y=0}^7 I(b_x,b_y)
\end{equation}
The range of the DC coefficient is [-1024,1016].

In step 4), a quantization matrix ${\bf Q}$ with 8$\times$8 components is used to obtain a matrix $\bf S_{q}$ from $\bf S$.
For example, 
\begin{equation}
\label{eq:quant}
S_{q} (u, v)=\mathrm{round}\left(\frac{S(u, v)}{Q(u, v)}\right),\ 0\leq u\leq 7,\  0\leq v \leq 7,
\end{equation}
where $S(u, v)$, $Q(u, v)$ and $S_{q}(u, v)$ represent the $(u,v)$ element of $\bf S$, $\bf Q$ and $\bf S_{q}$ respectively.
The $\mathrm{round}(x)$ function is used to round a value $x$ to the nearest integer value and $\lfloor x \rfloor$ denotes the integer part of $x$. 

The quality factor $QF\ (1\leq QF \leq 100)$ parameter is used to control a matrix $\bf Q$.
The large $QF$ results in a high quality image.

\subsection{Image Manipulation by SNS/CPSS Provider\label{sec:sn}}
Let us consider that JPEG images are uploaded to a SNS/CPSS provider.
It is known that JPEG images uploaded to SNS providers are often manipulated as below\cite{Intro,Intro1,Intro2}.
\begin{itemize}
 \item Editing metadata and filenames\\
Most of metadata in the header are deleted for privacy-concerns and the filenames of uploaded images are changed.
 \item Re-compressing uploaded images\\
Before stored in a cloud storage, uploaded images are decoded once and then the images are compressed again under the different coding condition.
 \item Resizing uploaded images\\
If uploaded images satisfy certain conditions, those images are resized.
For instance, in Twitter, when the filesize of images is larger than 3MB or the size of images is larger than 4096$\times$4096, the images will be resized.
\end{itemize}

As well as SNS providers, CPSS providers also manipulate uploaded images.
For instance, images uploaded to ``Google photos" are often re-compressed and resized.

In order to identify images uploaded to SNS/CPSS, it is required that the re-compression and resizing are considered in identification schemes, although conventional compression-dependent-schemes consider only the re-compression.

\subsection{Scenario}
\label{sec:id}
Let us consider a situation in which there are two or more compressed images generated under different or the same coding conditions.
They originated from the same image and were compressed under the various coding conditions.
We refer to the identification of those images as ``image identification".
Note that the aim of the image identification is not to retrieve visually similar images.

\begin{figure}[t]
\begin{center}
\begin{tabular}{c}
\begin{minipage}{\hsize}
  \begin{center}
   \includegraphics[width=85mm]{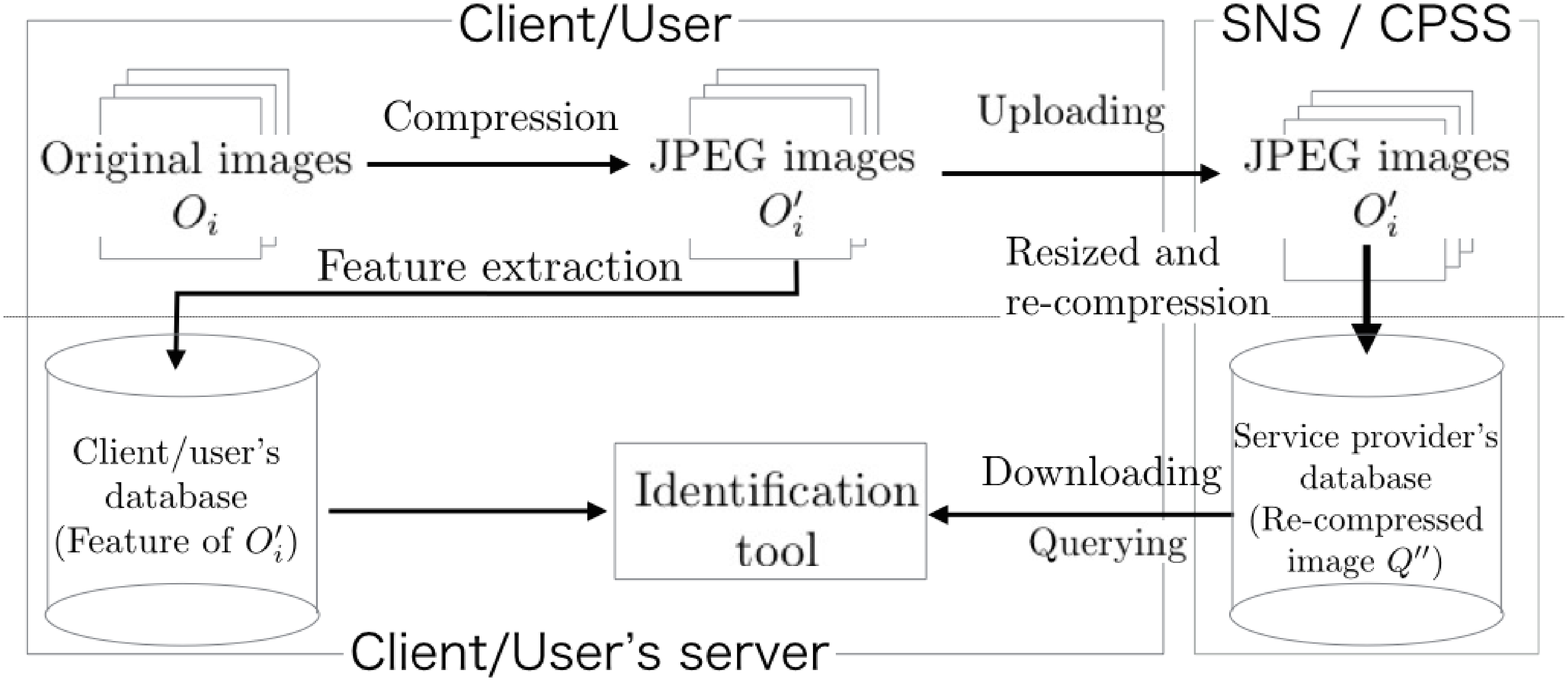}
      \end{center}
 \end{minipage}\\
\end{tabular}
\caption{Scenario}
 \label{fig:system}
 \end{center}
\end{figure} 

The scenario of this paper is illustrated in Fig. \ref{fig:system}.
In this scenario,  a client/user identifies images by using an identification tool.
When the client/user uploads JPEG images to SNS/CPSS, the features of these images are enrolled (extracted and then stored) in a client/user's database.
The uploaded images are resized to smaller sizes and re-compressed under different coding parameters, and then are stored in the cloud storage.
Finally, the client/user carries out the identification after extracting the feature from a query image i.e. a downloaded image.

\begin{figure}[t!] 
   \centering
   \includegraphics[width=8.5cm]{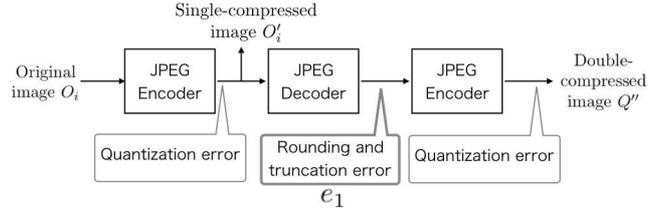} 
   \caption{JPEG errors in single-/double-compression}
   \label{fig:comp}
\end{figure}

The JPEG standard is generally used as a lossy compression method, so several errors are caused in the generation process of double-compressed images\cite{dj,zpid2}, as shown in Fig.\ref{fig:comp}.
In addition to ``quantization error" in the encoding process, ``rounding and truncation error" i.e. $e_1$ is caused in the decoding process.
In the proposed scheme, the errors in both processes are considered to identify double-compressed images.

\subsection{Notations and Terminologies}
The notations and terminologies used in the following sections are listed here.
\begin{itemize}
 \item
$O'_i$  represents a single-compressed image of an original image $O_i$.
 \item
$Q''$ represents a double-compressed query image. 
 \item
$M$ represents the number of 8$\times$8-blocks in an image. 
\item
$O'_i(m)$ and $q''(m)$ indicate quantized DC coefficients in $m$th block in images $O'_i$ and $Q''$ respectively ($0 \leq m < M$).
\item
$X_{O_i'}$ and $Y_{O_i'}$ represent the width and the height of $O_i'$ respectively.
As well, $X_{Q''}$ and $Y_{Q''}$ represent the width and the height of $Q''$ respectively.
\item
${Q_{O'_i,L}}$ and ${Q_{Q'',L}}$ indicate the DC components in the luminance quantization matrices, which are  used to generate images $O'_i$ and $Q''$ respectively.
 \item
$QF_{O'_i}$ and $QF_{Q''}$ indicate quality factors used to generate $O'_i$ and $\ Q''$ respectively. 
 \item
$\mathrm{sgn}(a)$ represents the sign of a real value $a$ as
\begin{equation}
\mathrm{sgn}(a)= \left \{
\begin{array}{c}
1,\  a>0,\\
0,\  a=0,\\
-1,\  a<0.\\
\end{array}
\right.
\end{equation}
\end{itemize}

\section{Proposed Identification Scheme\label{sec:proposed}}
The proposed identification scheme aims to identify double-compressed images.
In the proposed scheme, a feature of a JPEG image is extracted from only  DC coefficients of Y component.
Although the identification scheme explained in this section assumes the identification for the same size images, this scheme is easily extended for the identification of the different size images.
The enrollment and identification processes are performed as below.

\subsection*{1) Enrollment Process}
\begin{figure}[t!]
\begin{center}
\begin{tabular}{cc}
 \begin{minipage}{0.45\hsize}
  \begin{center}
   \includegraphics[width=40mm]{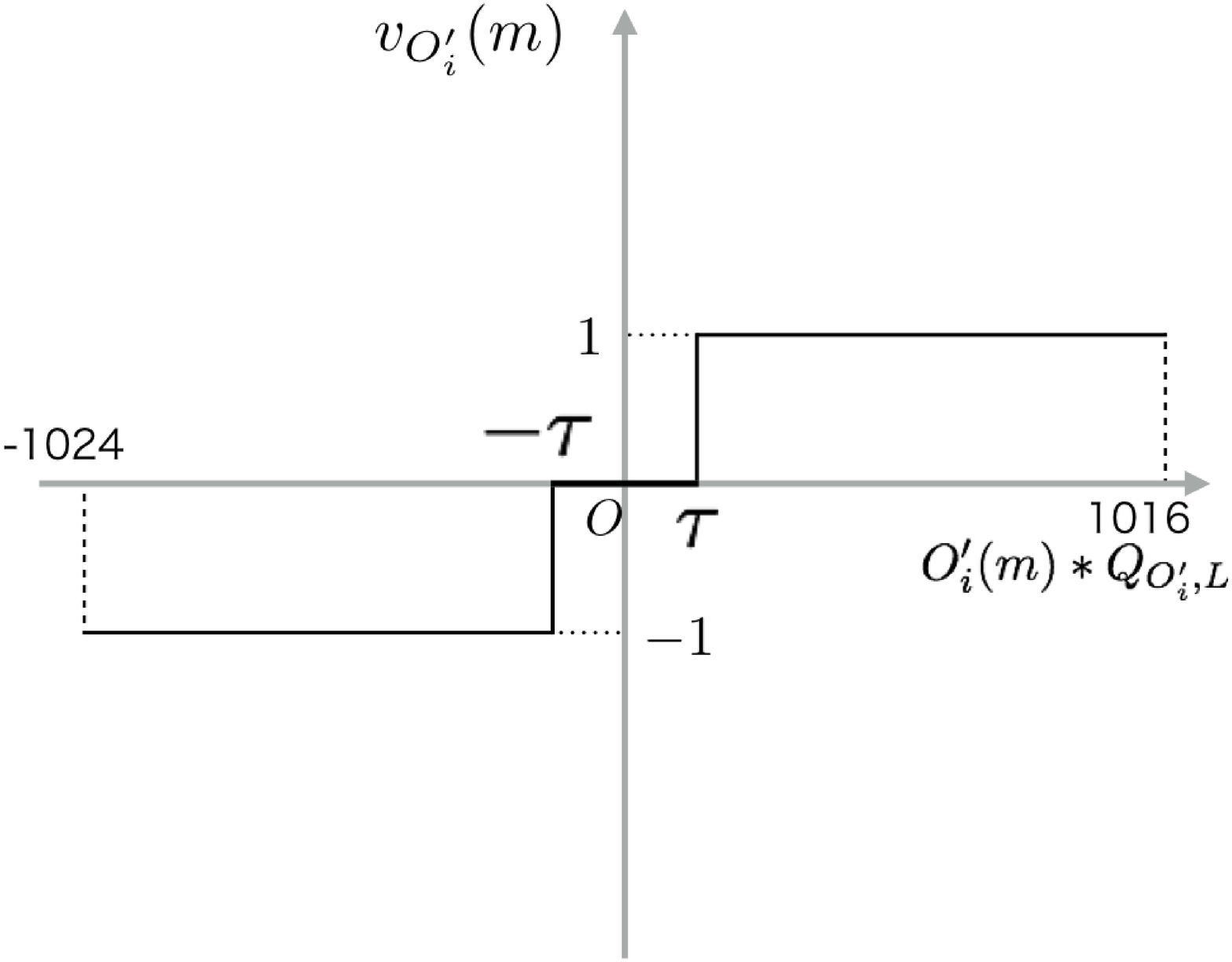}
  \end{center}
 \end{minipage}
  &\begin{minipage}{0.45\hsize}
  \begin{center}
   \includegraphics[width=40mm]{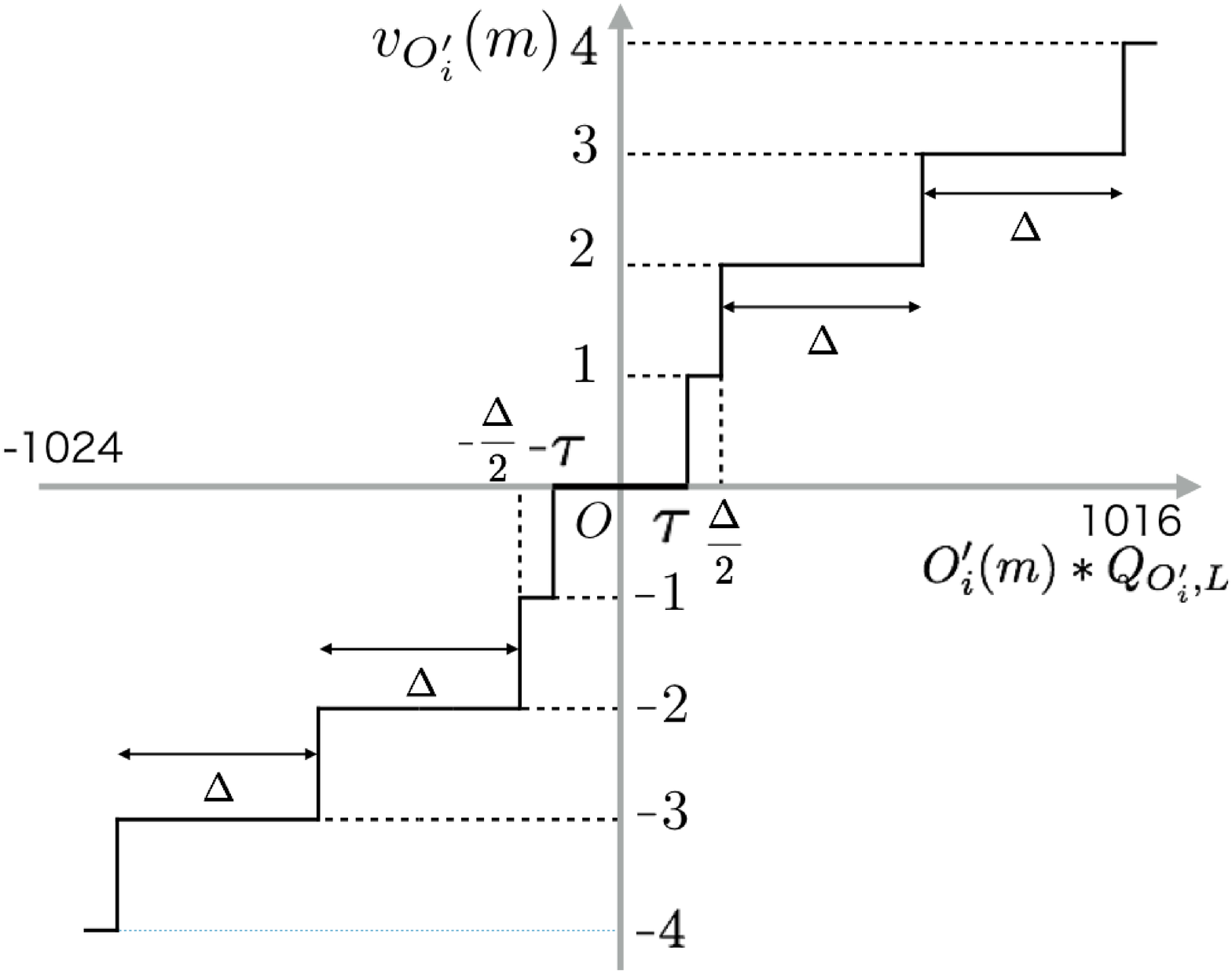}
  \end{center}
 \end{minipage}\\
(a) $\Delta >2048$ & (b) $\Delta \leq 2048$\\ 
 \end{tabular}
\caption{\textcolor{black}{Examples of the relation between $O_i'(m)$ and $v_{O_i'}(m)$, where $\tau=th*Q_{O_i',L}$}}
\label{fig:map}
 \end{center}
\end{figure}

\noindent In order to enroll image $O'_i$ as the feature vector ${\bf v}_{O_i'}\in \mathbb{R}^{\lceil \frac{X_{O_i'}}{8} \rceil * \lceil \frac{Y_{O_i'}}{8} \rceil \times 1}$,  a client/user carries out the following steps.
\begin{itemize}
 \item[(a)]
Set values $M$, $th$ and $\Delta$, where $th$ and $\Delta$ represent  a threshold value and a parameter used for the feature extraction.
 \item[(b)]
Set $m:=0$.
 \item[(c)]
Extract a component of the feature vector $v_{O_i'}(m)$ from a DC coefficient $O_i'(m)$ as  
\begin{equation}
\label{eq:map}
v_{O'_i}(m)= \left \{
\begin{array}{l}
0,\  -th  \leq O'_i(m) \leq th,\\
\begin{split}
\\&\mathrm{round}\left(\frac{Q_{O_i',L}*O_i'(m)}{\Delta}\right)\\ &+\mathrm{sgn( O_i'(m))} ,\, \mathrm{otherwise},
\end{split}
\end{array}
\right.
\end{equation}
where $v_{O'_i}(m)$ represents the $m$th component of the feature ${\bf v}_{O_i'}$. 
 \item[(d)]
Set $m:=m+1$.
If $m < M$, return to step (c). 
Otherwise, store ${\bf v}_{O'_i}$ as the feature  in the client/user's database.
\end{itemize}
For the feature extraction, a threshold value $th$ and a parameter $\Delta$ are used.
The aim of using $th$ is to avoid the effect of double-compression i.e. $e_1$, and $\Delta$ determines the amount of feature data stored in the database.
Figure \ref{fig:map} shows the relation between a DC coefficient $O_i'(m)$ and a component of the feature $v_{O_i'}(m)$.
As shown in Fig.\ref{fig:map}(a), when $\Delta>2048$, each component of the feature has one of three values, i.e., -1, 0, 1.
On the other hand,  the component of the feature generated with $\Delta \leq 2048$ has more various values (see in Fig.\ref{fig:map}(b)).
\textcolor{black}{As shown in Fig.\ref{fig:map}, $\tau=th*Q_{O_i',L}$ is a parameter to control robustness against errors caused by double-compression.}

\subsection*{2) Identification Process}
In order to compare image $Q''$ with image $O'_i$, the client/user carries out the following steps.
\begin{itemize}
 \item[(a)]
Set values $M$, $th$, $\Delta$, $d_{O_i'}$ and $d_{Q''}$, where $d_{O_i'}$ and $d_{Q''}$ are parameters for the identification.
It is required that the parameters $M$, $th$ and $\Delta$ are the same as those selected in step (a) of the enrollment process.
 \item[(b)]
Set $m:=0$. 
 \item[(c)]
Extract a component of the feature $v_{Q''}(m)$ from a DC coefficient $q''(m)$ as  
\begin{equation}
\label{eq:map}
v_{Q''}(m)= \left \{
\begin{array}{l}
0,\  -th  \leq q''(m) \leq th,\\
\begin{split}
\\&\mathrm{round}\left(\frac{Q_{Q'',L}*q''(m)}{\Delta}\right)\\ &+\mathrm{sgn( q''(m))} ,\, \mathrm{otherwise}.
\end{split}
\end{array}
\right.
\end{equation}
 \item[(d)]
If $|v_{O'_i}(m)|>d_{O_i'}$ or $|v_{Q''}(m)|>d_{Q''}$, proceed to step (f).
 \item[(e)]
If $\mathrm{sgn}(v_{O'_i}(m))\neq \mathrm{sgn}(v_{Q''}(m))$, the client/user judges that $O'_i$ and $Q''$ are generated from different original images and the process for image $O'_i$ is halted.
 \item[(f)]
Set $m:=m+1$.
If $m < M$, return to step (c).
Otherwise, the client/user judges that $O'_i$ and $Q''$ are generated from the same original image.
\end{itemize}
As shown above, by using the feature extracted from DC coefficients, the identification is carried out in the proposed scheme.
The following are the reasons why this feature is used in this paper.
\begin{itemize}
\item To identify different size images\\
In the case of using only DC coefficients mapped with $\Delta$, as shown in Sec. \ref{sec:modifyF}, DC coefficients in the resized image can be calculated from ones in the image before resizing.
\item To determine parameters independently of the size of images\\
The conventional scheme for double-compressed images \cite{zpid2}, which uses not only DC but also AC coefficients, requires the setting of the parameter related to the size of identified images. On the other hand, the parameters used in the proposed scheme, i.e. $th$ and $\Delta$ are independent of the size of images. 
\end{itemize}
In addition to these advantages, the use of $th$ allows us to reduce the influence of errors caused by the double-compression.

$d_{O_i'}$ and $d_{Q''}$ are required for the identification for different size images.

\section{Feature Modification for Identification of Different Size Images\label{sec:modifyF}}
In the processes mentioned above, the identification for the same size images is assumed.
However, images uploaded to SNS/CPSS providers are sometimes resized as smaller images.
Therefore, in order to identify different size images in the process mentioned in Sec. \ref{sec:proposed}, a modification method for the feature stored in the database is proposed in this section.
\subsection{Strategy for Resized Images} 
\begin{figure}[t!]
\begin{center}
\begin{tabular}{cc}
 \begin{minipage}{0.45\hsize}
  \begin{center}
   \includegraphics[width=40mm]{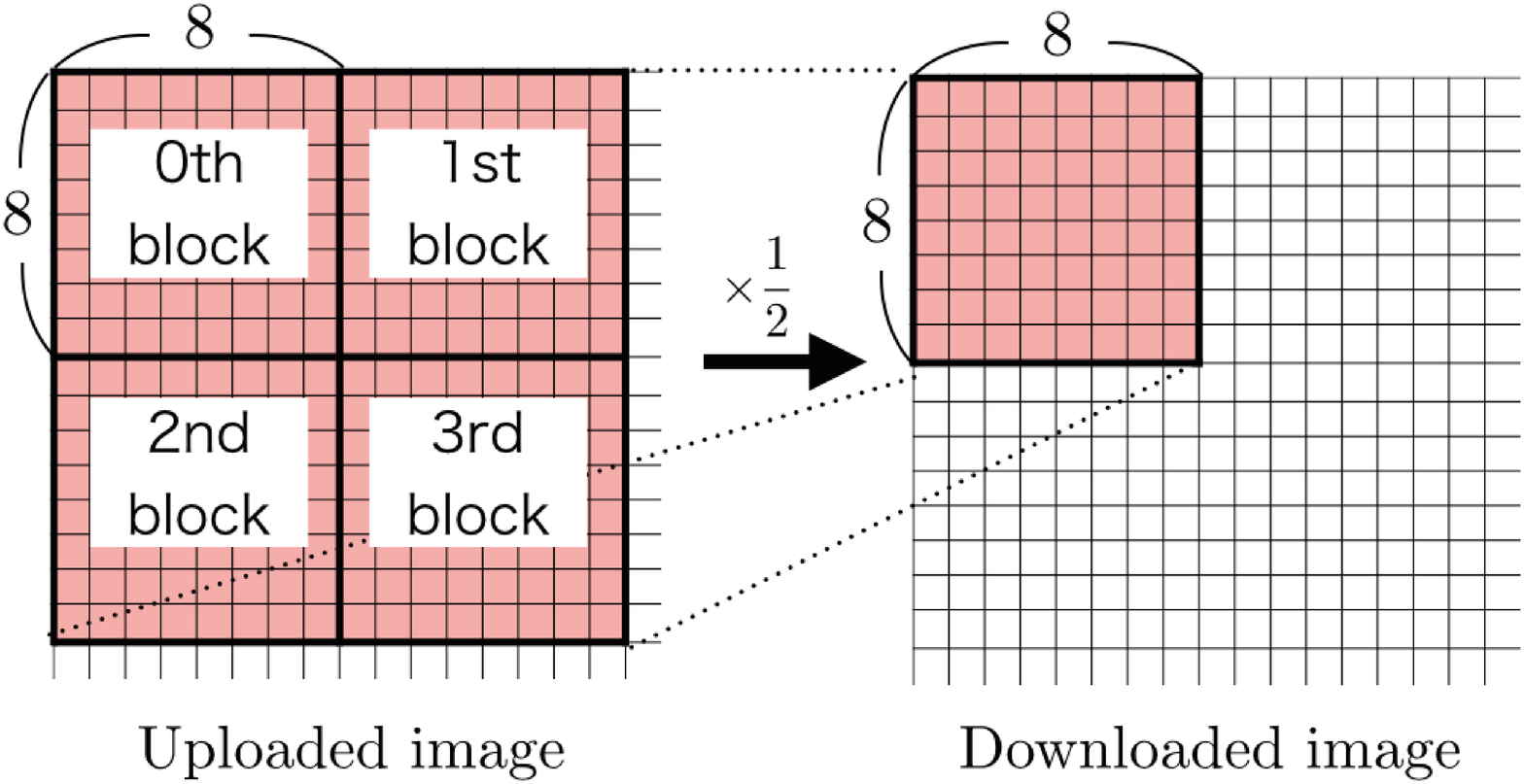}
  \end{center}
 \end{minipage}
  &\begin{minipage}{0.45\hsize}
  \begin{center}
   \includegraphics[width=40mm]{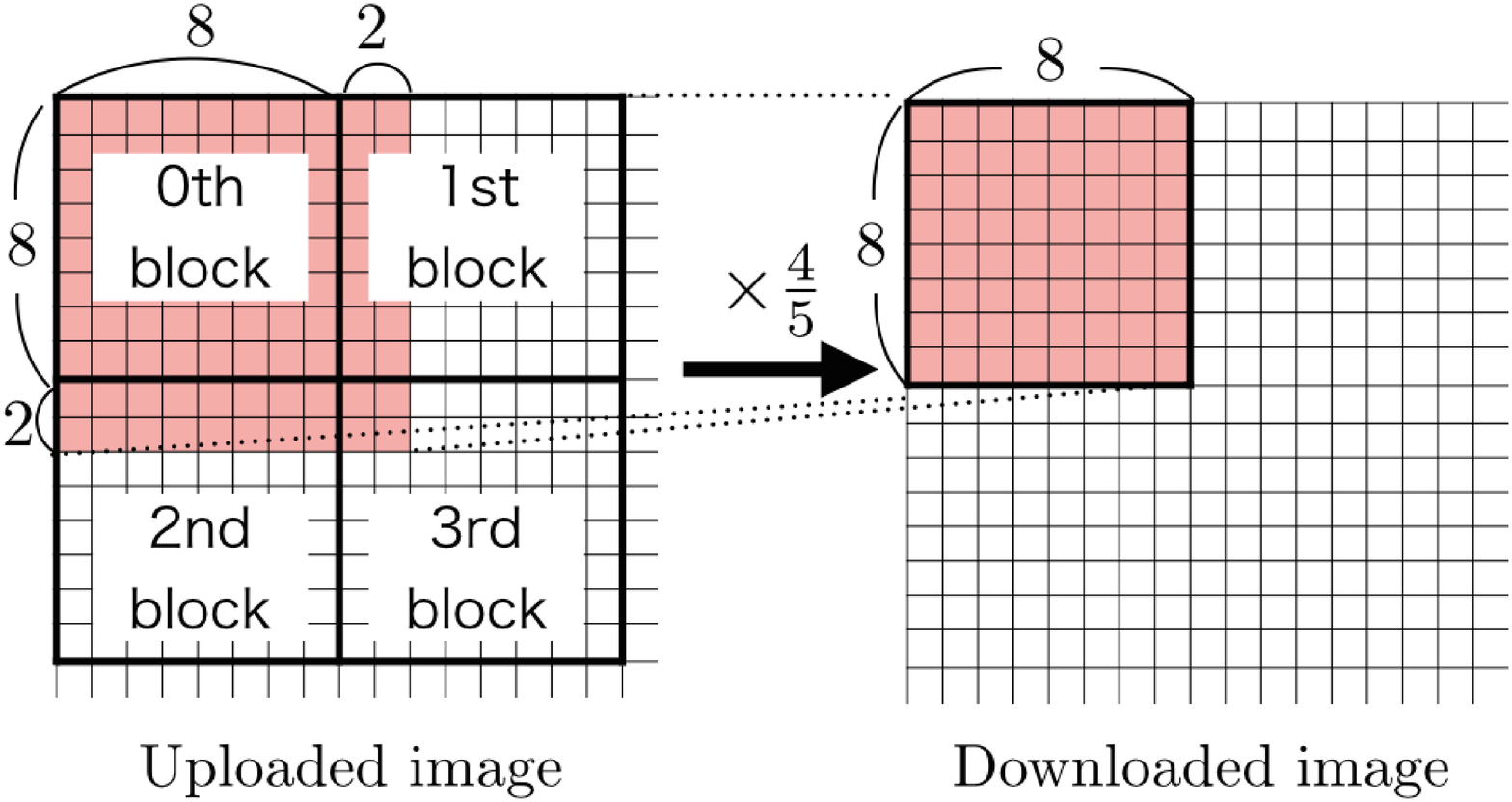}
  \end{center}
 \end{minipage}\\
(a) $s=2$ & (b) $s=\frac{5}{4}$\\ 
 \end{tabular}
\caption{Examples of the relationship between uploaded images and downloaded images with resizing}
\label{fig:exbl}
 \end{center}
\end{figure}

Let us consider that an uploaded image with the size of $Y\times X$ is resized to $\frac{1}{s}$ times size, i.e. $\lceil \frac{Y}{s} \rceil \times \lceil \frac{X}{s} \rceil$, where $s$ is a positive value and $\lceil a \rceil$ represents the ceiling of a real value $a$.
As shown in Fig.\ref{fig:exbl} (a), when the size of an uploaded image is changed to the half, i.e. $s=2$, 0th block in the downloaded image is computed by using four blocks from 0th block to 3rd block in the uploaded image.
The DC coefficient of every block $S(0,0)$ is defined by Eq.(\ref{eq:dc}), so DC coefficients in the downloaded image is estimated by calculating the average of the corresponding DC coefficients in the uploaded image as
\begin{equation}
q''(0)= \frac{1}{4} O_i'(0)+\frac{1}{4} O_i'(1)+\frac{1}{4} O_i'(2)+\frac{1}{4} O_i'(3).
\end{equation}

When $s$ is not an integer value as shown in Fig.\ref{fig:exbl} (b), the weighted average values of DC coefficients should be calculated, based on the number of corresponding pixels of each block in the uploaded image.  
For instance, as shown in Fig.\ref{fig:exbl} (b), i.e. for  $s=\frac{5}{4}$, the weights of four blocks are  $\frac{64}{100}, \frac{16}{100}, \frac{16}{100},\frac{4}{100}$ respectively.
Therefore, $q''(0)$ is estimated by 
\begin{equation}
\label{eq:est}
q''(0)=\frac{64}{100}O_i'(0)+ \frac{16}{100}O_i'(1)+ \frac{16}{100}O_i'(2)+\frac{4}{100}O_i'(3).
\end{equation}
Note that $10\times10=100$ pixels in an uploaded image are reduced to $8\times8=64$ pixels in the downloaded image for $s=\frac{5}{4}$.

In the practical, the estimated feature matrix $\bf D$ is computed by using the feature matrix $\bf U$ reproduced from the feature vector of an uploaded image ${\bf v}_{O_i'}$, where ${\bf U}  \in \mathbb{R}^{\lceil \frac{X_{O_i'}}{8} \rceil \times \lceil \frac{Y_{O_i'}}{8} \rceil}$ and ${\bf D}  \in \mathbb{R}^{\lceil \frac{X_{Q''}}{8} \rceil \times \lceil \frac{Y_{Q''}}{8} \rceil}$.
In the examples in Fig.\ref{fig:exbl}, the values in Eq.(\ref{eq:est}) can be expressed as
\begin{equation}
\label{eq:est2}
\begin{split}
D(0,0)=&\frac{64}{100}U(0,0)+ \frac{16}{100}U(1,0)\\ &+ \frac{16}{100}U(0,1)+\frac{4}{100}U(1,1),
\end{split}
\end{equation}
where $U(x_{O_i'},y_{O_i'})$ is the $(x_{O_i'},y_{O_i'})$ component of ${\bf U}$ $(0\leq x_{O_i'}<\lceil \frac{X_{O_i'}}{8} \rceil, 0\leq y_{O_i'} < \lceil \frac{Y_{O_i'}}{8} \rceil)$, and  ${\bf U}$ is mapped from ${\bf v}_{O_i'} $.

\subsection{Modification of Enrolled Feature}
According to  the strategy mentioned above, when the size of query images is not the same as that of the uploaded image, the enrolled features are modified before the identification process.
The modification process is shown as below.
\begin{itemize}
 \item[(a)]
Set values $X_{O_i'}$, $Y_{O_i'}$, $X_{Q''}$ and $Y_{Q''}$.
\item[(b)]
Map ${\bf v}_{O_i'}$ into a  matrix ${\bf U}$.
\item[(c)]
Set an estimated feature matrix ${\bf D}$ as a zero matrix.
\item[(d)]
Calculate $dx$ and $dy$ as below.
\begin{equation}
dx=\frac{8X_{O_i'}}{X_{Q''}},\  dy=\frac{8Y_{O_i'}}{Y_{Q''}}. 
\end{equation}
 \item[(e)]
Set $x:=0$, $y:=0$, $x_{Q''}:=0$ and $y_{Q''}:=0$.
\item[(f)]
Set $x_{O_i'}:=0$ and $y_{O_i'}:=0$.
\item[(g)]
Calculate a component of the weight matrix ${\bf W} \in \mathbb{R}^{\lceil \frac{X_{O_i'}}{8} \rceil \times \lceil \frac{Y_{O_i'}}{8} \rceil}$ by
\begin{equation}
W(x_{O_i'},y_{O_i'}) = \sum_{x_I=x_{O_i'}*8}^{x_{O_i'}*8+7} \sum_{y_I=y_{O_i'}*8}^{y_{O_i'}*8+7} \frac{Z(x_I,y_I)}{dx*dy} ,
\end{equation}
where $x_I$ and $y_I$ are integer values and 
\begin{equation}
Z(x_I,y_I) = \left \{
\begin{array}{ll}
\multirow{2}{*}{1,}& x \leq  x_I  <  x+dx  \\ 
 &\mathrm{and}\  y  \leq  y_I  <  y+dy,\\
0,& \mathrm{otherwise}.
\end{array}
\right.
\end{equation}
\item[(h)]
Update $D(x_{Q''},y_{Q''})$ by 
\begin{equation}
\begin{split}
D(x_{Q''},y_{Q''}) = D&(x_{Q''},y_{Q''})\\ &+W(x_{O_i'},y_{O_i'})*U(x_{O_i'},y_{O_i'}).
\end{split}
\end{equation}

\item[(i)]
Set $x_{O_i'}:= x_{O_i'} + 1$.
If $x_{O_i'} < \lceil \frac{X_{O_i'}}{8} \rceil$, return to step (g).
\item[(j)]
Set $x_{O_i'}:= 0$ and $y_{O_i'}:= y_{O_i'} + 1$.
If $y_{O_i'} <\lceil \frac{Y_{O_i'}}{8} \rceil$, return to step (g).
\item[(k)]
Set $x_{O_i'}:=0$, $y_{O_i'}:=0$, $x_{Q''}:= x_{Q''} + 1$ and $x:= x + dx$.
If $x+dx-1 < X_{O_i'}$, return to step (f).
\item[(l)]
Set $x_{Q''}:= 0$, $x:= 0$, $y_{Q''}:= y_{Q''} + 1$ and $y:= y + dy$.
If $y+dy-1 < Y_{O_i'}$, return to step (f).
\item[(m)]
Map  the estimated feature matrix $\bf D$ into a feature vector of the download image ${\bf v}_{D}\in \mathbb{R}^{\lceil \frac{X_{Q''}}{8} \rceil * \lceil \frac{Y_{Q''}}{8} \rceil\times 1}$.
 
\end{itemize}
After this modification, the identification process in Sec.\ref{sec:proposed} 2) can be carried out by replacing ${\bf v}_{O_i'}$ with ${\bf v}_{D}$.
Actually, there is the estimation error, so that $d_{O_i'}$ and $d_{Q''}$ are used for avoiding this error.

The use of the feature extracted from DC coefficients allows us not only to avoid the errors in double-compression but also to identify the different size images.
The effectiveness of the proposed scheme will be shown in Sec.\ref{sec:sim}.

\section{Simulation\label{sec:sim}}
A number of simulations were conducted to evaluate the performance of the proposed scheme.
We used the encoder and the decoder from the IJG (Independent JPEG Group) in the simulations\cite{ijg}.
\subsection{Selection of Threshold Value and Parameters}
\textcolor{black}{In order to select the values of $th$, $\Delta$, $d_{O_i'}$ and $d_{Q''}$, we conducted preliminary experiments as shown below.}
\color{black}
\subsection*{1). Determination of $th$}
\begin{itemize}
\item[(a)] Data set\\
885$\times$6 single-compressed images were generated from 885 original images in Uncompressed Color Image Database (UCID)\cite{ucid}  with six quality factors（$QF$ = 70, 75, 80, 85, 90, 95), and then every single-compressed one was re-compressed with six quality factors（$QF$ = 70, 75,80, 85, 90, 95) to obtain 885$\times$6$\times$6 double-compressed images.
\item[(b)]  Selection of single-compressed image\\
One single-compressed image was selected from 885$\times$6 single-compressed ones.
\item[(c)] Comparison of DC coefficients\\
At first, a double-compressed image was selected from six double-compressed images generated from the selected single-compressed one. Next, for all DC coefficients of two the selected JPEG images, the relation at the same block position between the two images was  investigated. When both DC coefficients in a block position have no zero value and the signs of the DC values are different,  a larger DC absolute value in the block was stored.  Accordingly, all larger DC ones at the blocks at which the above condition was satisfied were stored. This process was conducted for six corresponding double-compressed images.
\end{itemize}
Step (b) and step (c) were carried out until all single-compressed images were selected in step (b).
\begin{itemize}
\item[4)] Selection of $th$\\
The largest absolute value in the stored ones was chosen as $th$.
\end{itemize}
According to the above procedure, $th$ was experimentally determined as 14. The parameter $th$ is used to skip small DC coefficients at step (d) of the identification process, because the signs of such coefficients are easily inverted by the effect of double-compression.

\subsection*{2). Determination of $\Delta$}
After step (a) and step (b) as mentioned above, the following steps were conducted.
\begin{itemize}
\item[(c)] Calculate the differences between DC coefficients \\
At first, a double-compressed image was selected from six double-compressed images generated from the selected single-compressed one. Next, for all DC coefficients of two the selected images, the relation of DC coefficients at each position was investigated. When both DC coefficients at each position had the same sign and larger absolute values than $th$,  the absolute value of the difference between two the DC values was saved for all positions at which the condition was satisfied, respectively. This process was conducted for six corresponding double-compressed images.

\end{itemize}
Step (b) and step (c) were carried out until the all single-compressed images were selected in step (b).
\begin{itemize}
\item[(d)] Selection of $\Delta$\\
The largest absolute value in the stored ones was selected as $\Delta$.

\end{itemize}
By conducting the above steps, $\Delta$ was determined as 50. 

 \subsection*{3). Determination of $d_{O_i'}$ and $d_{Q''}$}
Using $th$ and $\Delta$ selected by the above procedures, the identification experiments were performed by using various JPEG images while changing the values of $d_{O_i'}$ and $d_{Q''}$. 
From the result,  $d_{O_i'}$ and $d_{Q''}$ were determined as 4.

As shown in simulation results later, the use of the parameters $th = 14$, $\Delta=50$, $d_{O_i'}=4$ and $d_{Q''}=4$ provided a high performance, so this selection was good one, although other selections provided almost the same results. 
\color{black}
\subsection{Querying Performance}
\begin{figure}[t!]
\begin{center}
\begin{tabular}{c}
 \begin{minipage}{0.3\hsize}
  \begin{center}
   \includegraphics[width=23mm]{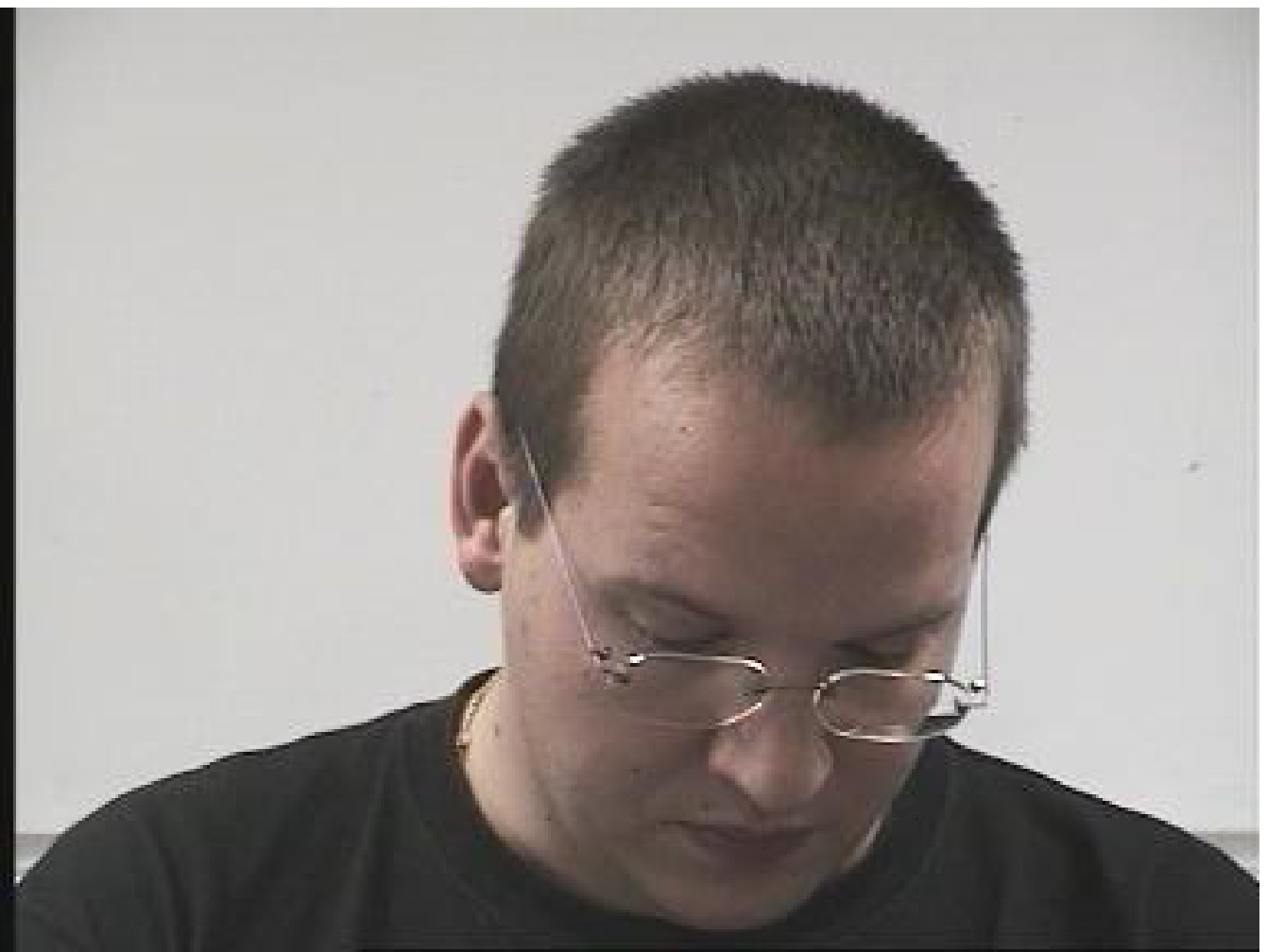}
  \end{center}
 \end{minipage}
  \begin{minipage}{0.3\hsize}
  \begin{center}
   \includegraphics[width=23mm]{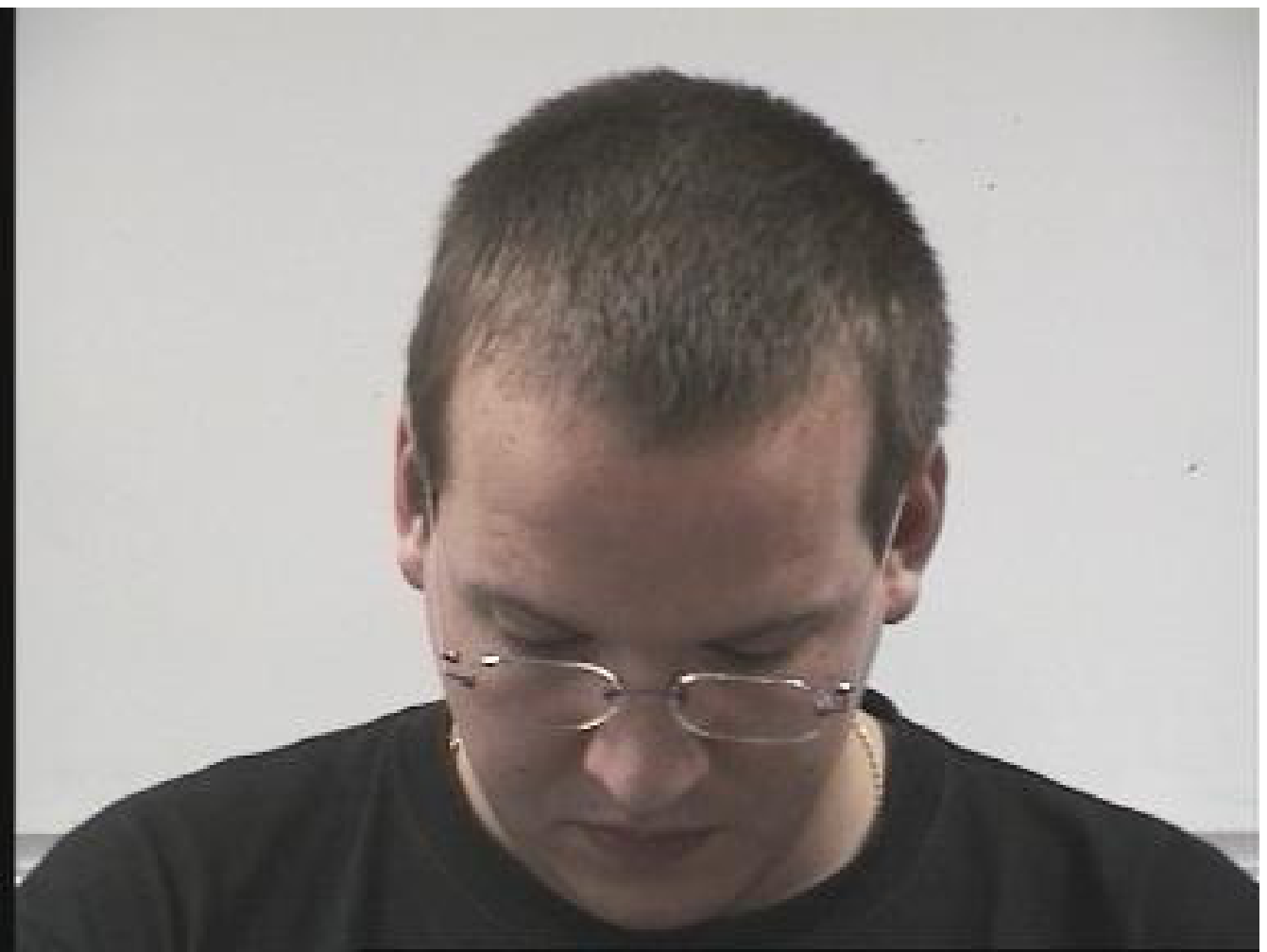}
  \end{center}
 \end{minipage}
  \begin{minipage}{0.3\hsize}
  \begin{center}
   \includegraphics[width=23mm]{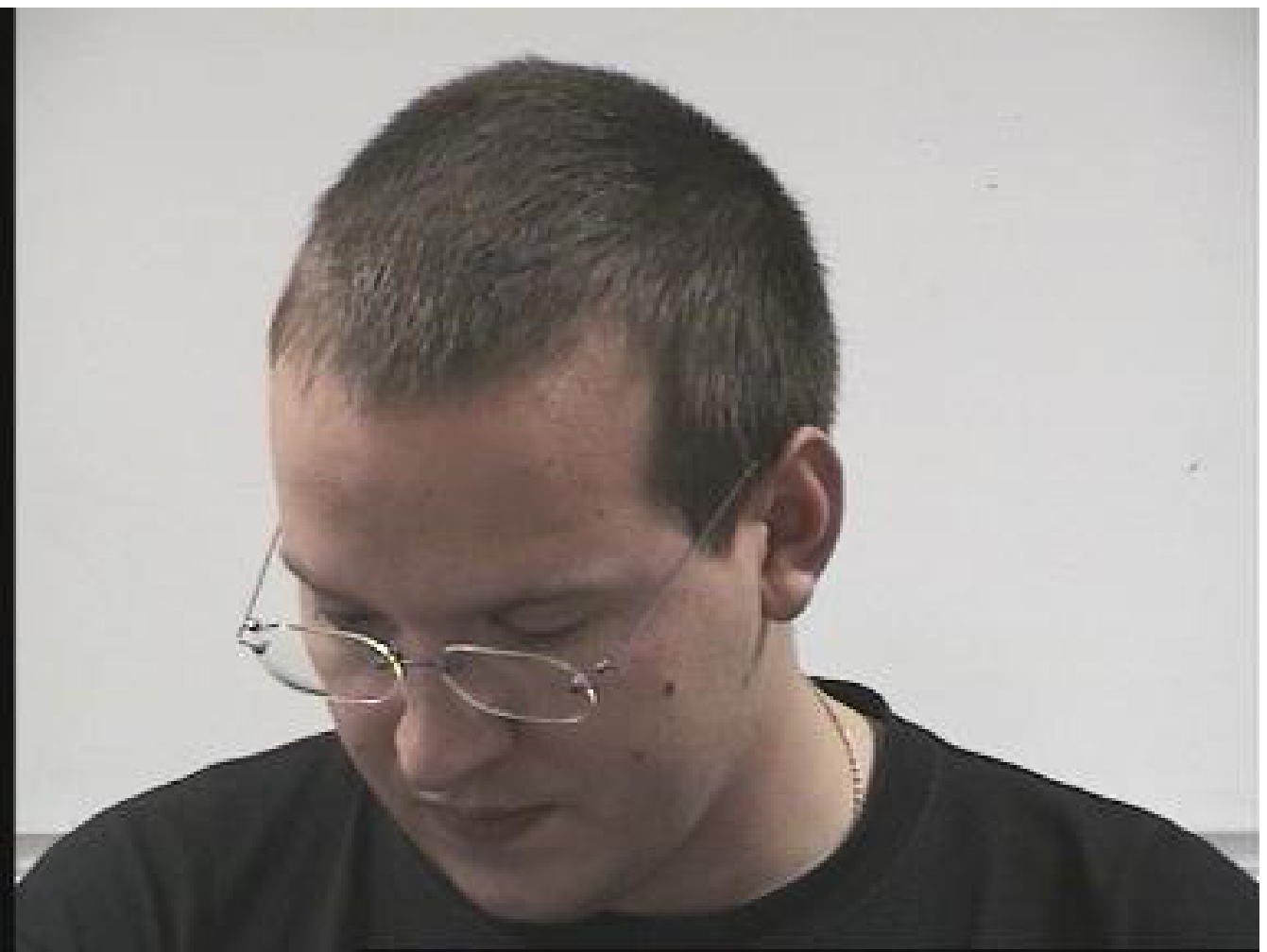}
  \end{center}
 \end{minipage}
 \end{tabular}
\caption{Examples of test images in HPID (288$\times$384)}
\label{fig:exhpid}
 \end{center}
\end{figure}

Next, we used the images in Head Pose Image Database (HPID) \cite{hpid} to evaluate querying performance.
As shown in Fig. \ref{fig:exhpid}, HPID consists of very similar images, where the size of images is 288$\times$384. 
The main reason of using HPID is to show that the proposed scheme can detect a slight differences between the images.
Therefore, we used 186 images of ``Person01'' in HPID as original images.

The proposed scheme was compared with two compression-method-dependent-schemes (zero positions-based scheme \cite{zpid2} and DC signs-based scheme\cite{dc}) and three image hashing-based schemes (low-rank and sparse decomposition-based scheme\cite{ih1}, quaternion-based one \cite{ih2} and iterative quantization (ITQ)-based one\cite{itq}), where ITQ-based hash values were generated from 512 dimensional GIST feature vectors and each hash value was represented by 512 bits.  
In the schemes\cite{ih1,ih2,itq}, the hamming distances between the hash value of a query image and those of all images in each database were calculated, and then images that had the smallest distance were chosen as the images generated from the same original image as the query, after decompressing all images.

\subsection*{1)Querying Performance for Images without Resizing}
At first, the querying performances for images without resizing were evaluated.
Table \ref{tab:condition} summarizes the quality factors used to generate JPEG images, where $DB_1$, $DB_2$ and $DB_3$ indicate the databases of client/user in Fig.\ref{fig:system}.
\textcolor{black}{First of all, 186 single-compressed images were generated from 186 original images for each database, so  558 single-compressed images were generated from original ones for three databases. 
Next, those single-compressed images were re-compressed with four quality factors i.e. $QF_{Q''}=71, 75,80,85$, where 744 double-compressed images were generated for each database. 
It is known that the range of quality factors used for re-compression in SNS  is [71,85] as in \cite{Intro2}, so these quality factors were used.
Thus, to confirm whether each query image has the same original image as one of 186 single-compressed images, 186$\times$744 identification operations were carried out for each database.
For instance, in order to perform the identification operations for $DB_1$, features were extracted from 186 single-compressed images compressed with $QF_{O'_i}=95$, and 186 images with $QF_{O'_i}=95$ were re-compressed with $QF_{Q''}=71,75,80,85$ respectively to generate 744 query images for $DB_1$.}
\begin{table}[t!]
\caption{Quality factors used to generate JPEG images\label{tab:condition}.$DB_1$, $DB_2$ and $DB_3$ indicate databases of client/user in Fig.\ref{fig:system}}
\centering
\scalebox{.9}{
\begin{tabular}{|c|c|c|c|c|c|}\hline
\multicolumn{2}{|c|}{JPEG images}&Quality factors\\\hline
\multirow{3}{*}{\shortstack{Uploaded images }}&$DB_1$&$QF_{O'_i}=95$\\\cline{2-3}
&$DB_2$&$QF_{O'_i}=85$\\\cline{2-3}
&$DB_3$&$QF_{O'_i}=75$\\\hline
\multicolumn{2}{|c|}{\shortstack{Downloaded images \\(Query images $Q''$)}} &$QF_{Q''}=71,75,80,85$\\\hline
\end{tabular}
}\end{table}

Table \ref{tab:res2}  shows $Precision$ and $Recall$,  defined by
\begin{equation}
Precision = \frac{TP}{TP+FP},\ Recall = \frac{TP}{TP+FN},
\end{equation}
where TP, FP and FN represent the number of true positive, false positive and false negative matches respectively.
Note that $Recall=100[\%]$ means that there were no false negative matches, and $Precision=100[\%]$ means that there were no false positive matches.

It is confirmed that the proposed scheme and two-compression-method-dependent ones achieved both $Recall=100[\%]$ and $Precision=100[\%]$, although the image hashing-based ones did not.

\begin{table}[t!]
\caption{Querying performances for images in HPID}
\label{tab:res2}
\centering
\scalebox{.9}{
\begin{tabular}{|c|c|c|c|c|c|c|c|c|}\hline
\multirow{1}{*}{scheme}&\multirow{1}{*}{database}&$Precision$[\%]&$Recall$[\%]\\\hline
\multirow{3}{*}{\shortstack{proposed\\($\Delta=50,d_{O_i'}=4, d_{Q''}=4$)}}&$DB_1$&100&100\\\cline{2-4}
&$DB_2$&100&100\\\cline{2-4}
&$DB_3$&100&100\\\hline
\multirow{3}{*}{\shortstack{DC signs\cite{dc}}}&$DB_1$&100&100\\\cline{2-4}
&$DB_2$&100&100\\\cline{2-4}
&$DB_3$&100&100\\\hline
\multirow{3}{*}{\shortstack{zero value\\ positions\cite{zpid2}}}&$DB_1$&100&100\\\cline{2-4}
&$DB_2$&100&100\\\cline{2-4}
&$DB_3$&100&100\\\hline
\multirow{3}{*}{\shortstack{low-rank\\and sparse\\decomposition\cite{ih1}}}&$DB_1$&97.21&98.39\\\cline{2-4}
&$DB_2$&98.41&99.73\\\cline{2-4}
&$DB_3$&96.35&99.33\\\hline
\multirow{3}{*}{\shortstack{quaternion\cite{ih2}}}&$DB_1$&99.60&100\\\cline{2-4}
&$DB_2$&99.60&99.87\\\cline{2-4}
&$DB_3$&100&100\\\hline
\multirow{3}{*}{\shortstack{ITQ\cite{itq}}}&$DB_1$&67.24&99.33\\\cline{2-4}
&$DB_2$&67.67&99.87\\\cline{2-4}
&$DB_3$&62.98&98.79\\\hline
\end{tabular}}
\end{table}

\subsection*{2)Querying Performances for Images with Resizing}
Next,  the querying performances for images with resizing were evaluated.
The images stored as the feature and query images were generated by following the conditions shown in Tab. \ref{tab:conditionResize}.
For instance, after 186 original images were resized to the size 960$\times$1280 and compressed with $QF_{O'_i}=95$, features stored in the database $DB_4$ were extracted from the generated images.
In order to  generate query images for $DB_4$, the images enrolled as features were resized to the size 720$\times$960, and then the resized images were compressed with  $QF_{Q''}=71,75,80,85$.

Table \ref{tab:res}  shows the results, where ``-" means that the identification can not be applied.
The two compression-method-dependent schemes \cite{zpid2,dc} assume the identification for the same size images, so that they were not evaluated in this simulation. 
It is confirmed from Table \ref{tab:res} that only the querying performances of the proposed scheme were perfect as well as the performances for the same size images.
Therefore, the proposed scheme outperformed the conventional schemes as well as for images without resizing.

\begin{table}[t!]
\caption{Sizes of resized images and quality factors used to generate JPEG images\label{tab:conditionResize}.$DB_4$, $DB_5$, $DB_6$, $DB_7$, $DB_8$ and $DB_9$ indicate databases of client/user in Fig.\ref{fig:system}}
\centering
\scalebox{1}{
\begin{tabular}{|c|c|c|c|c|c|}\hline
\multicolumn{2}{|c|}{JPEG images}&Size&Quality factors\\\hline
\multirow{6}{*}{\shortstack{Uploaded images}}&$DB_4$&960$\times$1280&$QF_{O'_i}=95$\\\cline{2-4}
&$DB_5$&960$\times$1280&$QF_{O'_i}=85$\\\cline{2-4}
&$DB_6$&960$\times$1280&$QF_{O'_i}=75$\\\cline{2-4}
&$DB_7$&1440$\times$1920&$QF_{O'_i}=95$\\\cline{2-4}
&$DB_8$&1440$\times$1920&$QF_{O'_i}=85$\\\cline{2-4}
&$DB_9$&1440$\times$1920&$QF_{O'_i}=75$\\\hline
\multicolumn{2}{|c|}{\shortstack{Downloaded images \\(Query images $Q''$)}} &720$\times$960&$QF_{Q''}=71,75,80,85$\\\hline
\end{tabular}
}\end{table}

\begin{table}[t!]
\caption{Querying performance for resized images, where ``-" means that the identification can not be applied.}
\label{tab:res}
\centering
\scalebox{.9}{
\begin{tabular}{|c|c|c|c|c|c|c|c|c|}\hline
\multirow{1}{*}{scheme}&\multirow{1}{*}{database}&$Precision$[\%]&$Recall$[\%]\\\hline
\multirow{6}{*}{\shortstack{proposed\\($\Delta=50,d_{O_i'}=4, d_{Q''}=4$)}}&$DB_4$&100&100\\\cline{2-4}
&$DB_5$&100&100\\\cline{2-4}
&$DB_6$&100&100\\\cline{2-4}
&$DB_7$&100&100\\\cline{2-4}
&$DB_8$&100&100\\\cline{2-4}
&$DB_9$&100&100\\\hline
\multirow{6}{*}{\shortstack{DC Signs\cite{dc}}}&$DB_4$&-&-\\\cline{2-4}
&$DB_5$&-&-\\\cline{2-4}
&$DB_6$&-&-\\\cline{2-4}
&$DB_7$&-&-\\\cline{2-4}
&$DB_8$&-&-\\\cline{2-4}
&$DB_9$&-&-\\\hline
\multirow{6}{*}{\shortstack{zero value\\ positions\cite{zpid2}}}&$DB_4$&-&-\\\cline{2-4}
&$DB_5$&-&-\\\cline{2-4}
&$DB_6$&-&-\\\cline{2-4}
&$DB_7$&-&-\\\cline{2-4}
&$DB_8$&-&-\\\cline{2-4}
&$DB_9$&-&-\\\hline

\multirow{6}{*}{\shortstack{low-rank\\and sparse\\decomposition\cite{ih1}}}&$DB_4$&97.37&99.60\\\cline{2-4}
&$DB_5$&97.10&98.92\\\cline{2-4}
&$DB_6$&95.50&99.73\\\cline{2-4}
&$DB_7$&96.86&99.60\\\cline{2-4}
&$DB_8$&97.49&99.19\\\cline{2-4}
&$DB_9$&97.48&98.79\\\hline
\multirow{6}{*}{\shortstack{quaternion\cite{ih2}}}&$DB_4$&98.94&100\\\cline{2-4}
&$DB_5$&99.73&99.73\\\cline{2-4}
&$DB_6$&98.80&100\\\cline{2-4}
&$DB_7$&99.73&99.73\\\cline{2-4}
&$DB_8$&98.94&100\\\cline{2-4}
&$DB_9$&98.94&100\\\hline
\multirow{6}{*}{\shortstack{ITQ\cite{itq}}}&$DB_4$&72.65&94.62\\\cline{2-4}
&$DB_5$&77.58&97.17\\\cline{2-4}
&$DB_6$&75.53&96.64\\\cline{2-4}
&$DB_7$&52.58&79.44\\\cline{2-4}
&$DB_8$&54.05&78.09\\\cline{2-4}
&$DB_9$&50.99&79.30\\\hline
\end{tabular}}
\end{table}

\section{Conclusion}
In this paper, a new image identification scheme for double-compressed JPEG images was proposed to relate a query image with images uploaded to SNS/CPSS.
The proposed scheme uses a feature extracted from DC coefficients in Y component.
The use of the feature allows us to avoid the errors caused by double-compression. 
In addition, the identification for the different size images can be performed, although the conventional compression-method-dependent schemes can not.
The simulation results showed that the proposed scheme detected slightly differences and outperformed  other schemes including the state-of-art one, even if images were very similar.
We plan to extend the proposed scheme as a tamper localization in our future work. 

\bibliography{ref}
\end{document}